\documentclass[aps,prb,twocolumn,groupedaddress,showpacs]{revtex4}

\usepackage{epsfig}

\begin{document}

\title{Determination of the Carrier-Envelope Phase of Few-Cycle Laser
Pulses with Terahertz-Emission Spectroscopy}

\author{Markus Kre\ss$^1$, Torsten L\"offler$^1$, Mark D. Thomson$^1$, Reinhard D\"orner$^2$,
Hartmut Gimpel$^3$, Karl Zrost$^3$, Thorsten Ergler$^3$, Robert
Moshammer$^3$, Uwe Morgner$^{3,4}$, Joachim Ullrich$^3$, and
Hartmut G. Roskos$^1$}

\affiliation{$^1$ Physikalisches Institut,
Johann~Wolfgang~Goethe--Universit\"at, Max-von-Laue-Str.~1,
D--60438 Frankfurt (M), Germany} \affiliation{$^2$ Institut f\"ur
Kernphysik, Johann~Wolfgang~Goethe--Universit\"at,
Max-von-Laue-Str.~1, D--60438 Frankfurt (M), Germany}
\affiliation{$^3$ Max-Planck-Institut f\"ur Kernphysik,
Saupfercheckweg~1, D--69117 Heidelberg, Germany} \affiliation{$^4$
Institut f\"ur Quantenoptik, D-30167 Hannover, Germany}
\affiliation{Correspondence should be addressed to - [MK]}


\date{March 16, 2006}

\maketitle
\begin{bf}

The availability of few-cycle optical pulses opens a window to
physical phenomena occurring on the attosecond time scale. In
order to take full advantage of such pulses, it is crucial to
measure \cite{Baltuska,baltuska2,Apolonski,Paulus} and stabilise
\cite{Baltuska,baltuska2} their carrier-envelope (CE) phase, i.e.,
the phase difference between the carrier wave and the envelope
function. We introduce a novel approach to determine the CE phase
by down-conversion of the laser light to the terahertz (THz)
frequency range via plasma generation in ambient air, an isotropic
medium where optical rectification (down-conversion) in the
forward direction is only possible if the inversion symmetry is
broken by electrical or optical means
\cite{Löffler,Löffler3,Cook2,Kress,Reimann,Zhang}. We show that
few-cycle pulses directly produce a spatial charge asymmetry in
the plasma. The asymmetry, associated with THz emission, depends
on the CE phase, which allows for a determination of the phase by
measurement of the amplitude and polarity of the THz pulse.
\end{bf}

The ability to measure \cite{Baltuska,baltuska2,Apolonski,Paulus}
and stabilise \cite{Baltuska,baltuska2} the CE phase, which is
also often named absolute phase and defined by Eq.\,\ref{pulse}
(Methods), of few-cycle laser pulses, i.e., the ability to
completely control ultrashort high-intensity electromagnetic
fields, provides access to a whole series of novel physical
phenomena, and creates options for their coherent control. An
example is the manipulation of the electron recombination process
during ``re-collision'' \cite{Itatani}, with the aim to fully
control the generation of both intense VUV higher-harmonic
radiation \cite{Macklin,Gordienko} as well as of high-intensity
single attosecond pulses. The progress has led to the
establishment of a new field of research, ``attosecond science'',
with highlights of recent work including the efficient tomography
of molecular wave functions by coherent diffraction of rescattered
electrons \cite{Itatani2} as well as the preparation of
high-intensity, monochromatic femtosecond electron beams evolving
from ``bubble generation'' \cite{Pukhov}. Many of the phenomena
studied depend critically on the CE phase. Examples include the
dynamics of above-threshold ionisation (ATI) \cite{Paulus},
light-induced non-sequential double ionisation \cite{Rottke}, Gouy
phase observation \cite{Lindner}, and quantum interference in
photocurrents \cite{Roos05}. An exciting prospect is the
application of the CE phase as a coherent-control tool, e.g., to
steer bound electrons in molecules \cite{Bandrauk}.

The state-of-the-art method for CE phase determination of
amplified laser pulses is stereo ATI \cite{paulus3,paulus2}. In
this letter, we present a new approach based on down-conversion
into the THz frequency range in a laser-generated ambient
air-plasma. The concept relies on the detection of the THz
electromagnetic pulses emitted during photo-ionization of the air
by the focussed few-cycle pulses. The method is easy to apply,
because it does not require a vacuum chamber and relies on
standard optoelectronic THz detection
technology\,\cite{Löffler,Kress}.

\begin{figure}
\begin{center}
\epsfbox{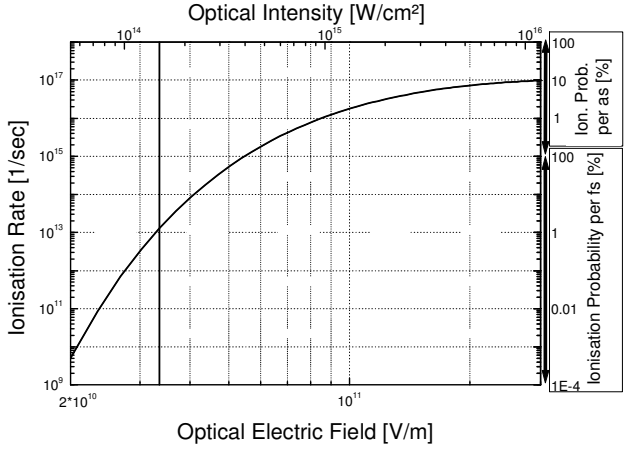} \caption{Ionisation rate for $N_2$ vs.
electric field strength based on the static-tunnelling theory as a
function of the optical field strength, respectively the light
intensity. The calculation was performed with the help of
Eq.\,\ref{eqn1} of Sec.\,\ref{meth}. An ionisation rate of
$10^{15}$\,s$^{-1}$ corresponds to a probability of 100 $\%$ for a
molecule to be ionised during a time interval of 1 fs. Note, that
the probability for ionisation of a molecule has a value of $1\%$
per fs at a threshold field of 3.3$\cdot$10$^{10}$\,V/m (vertical
black line), and that the ionisation probability per fs already
becomes 100\% at 5.5$\cdot$10$^{10}$\,V/m, i.e., at less than
twice the threshold field.} \label{ion-rate-a}

\end{center}
\end{figure}

\begin{figure}
\begin{center}
\epsfbox{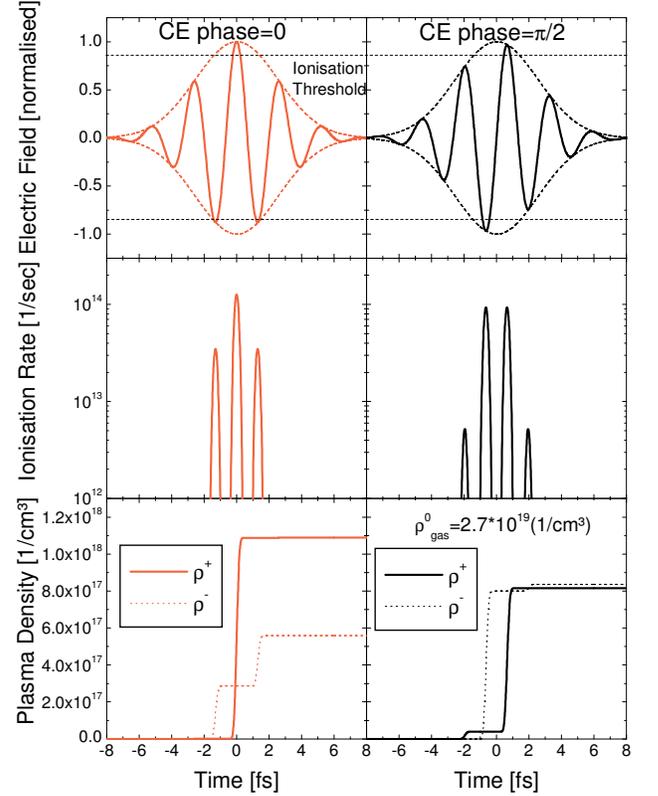} \caption{Optical electric field, ionisation
rate and plasma densities as a function of time for CE phases
$\varphi_{CE}=0$ and $\varphi_{CE}=\pi/2$. The plasma densities
$\rho^+$ and $\rho^-$, generated at positive respectively negative
field polarities, are calculated with Eqs.\,\ref{diff_ion-1} and
\ref{diff_ion-2} of Sec.\,\ref{meth}. The ionisation threshold is
the arbitrarily chosen electric-field value where the ionisation
probability is $1\%$ per fs.} \label{ion-rate-b}
\end{center}
\end{figure}

THz-pulse emission from laser-generated plasmas has been
investigated in the past, but only with much longer laser pulses.
There, the transient electric currents responsible for the THz
radiation have their origin either in ponderomotive forces
\cite{Hamster}, the Coulomb force of an externally applied DC bias
\cite{Löffler,Löffler3}, or an optical ``AC bias'' achieved by
superimposing the light pulses with their own second-harmonic
radiation \cite{Cook2,Kress,Reimann,Zhang}. The THz-wave
generation mechanism for the latter has been discussed in terms of
a four-wave-mixing (FWM) process\,\cite{Cook2,Reimann,Zhang}. For
the present case where the pulse spectra are octave-spanning FWM
results in a low-frequency ($\Omega\rightarrow0$) polarisation:
$P\left( {\Omega  = \omega _1  + \omega _2  - \omega _3 } \right)
\propto \sqrt {I\left( {\omega _1 } \right)I\left( {\omega _2 }
\right)I\left( {\omega _3 } \right)} \cos \left( \varphi_{CE}
\right)$, because the CE phases $\varphi_{CE}$ of all three input
waves are the same (as all frequency components arise from the
same near-transform-limited few-cycle pulse). The CE phase is
hence directly imparted onto the single-cycle THz pulse and can
thus be determined via electrooptic measurement of the THz field
amplitude. The same argument concerning the transfer of the CE
phase to the THz field can be readily generalised to the case of
higher-order non-linear
contributions.\\
Whilst such a wave-mixing process illustrates the mechanism for CE
phase-sensitive THz generation, we will give an alternative
description based on static-tunnelling theory\cite{Corkum89},
which explains the origin of non-linear polarisation on a
microscopic level. For simplicity, we assume the air to consist
only of N$_2$. Figure~\ref{ion-rate-a} displays the calculated
ionisation rate $w_{tun}$ for N$_2$ at a pressure of 1 atm,
respectively the ionisation probability rate of a single molecule.
The strongly non-linear intensity dependence of the ionisation
rate translates into temporal ionisation patterns as illustrated
in Fig.\,\ref{ion-rate-b}. The upper panels display the two pulse
shapes considered here, both with the same pulse duration of
$\tau=6$\,fs (FWHM), the same centre wavelength of
$\lambda=800$\,nm, and the same peak electric field of the
envelope of $E^0_{opt}=4.2 \cdot$10$^{10}$\,V/m, but with
different CE phases, $\varphi_{CE}=0$ and $\varphi_{CE}=\pi/2$.
The middle panels of Fig.\,\ref{ion-rate-b} exhibit the time
evolution of the ionisation rates for the two cases. A common
signature is that only the three or four most intense half-cycles
of the pulses ionise a significant amount of molecules. The
ionisation rate at each half-cycle, however, depends sensitively
on the CE phase. This becomes even clearer when we calculate the
temporal evolution of the plasma density. We distinguish two
quantities, $\rho^+(t)$ and $\rho^-(t)$, which describe the
densities of free electrons produced at positive and negative
polarity of the optical laser field, respectively. The total
density of free electrons is given by $\rho_{total} = \rho^+(t) +
\rho^-(t)$. The lower pair of panels of Fig.\,\ref{ion-rate-b}
shows that for $\varphi_{CE}=0$, the total plasma density
generated at negative field polarity is significantly smaller than
the density for positive polarity. For $\varphi_{CE}=\pi/2$, the
densities only differ little after the pulse has passed through
the focus.

\begin{figure}
\begin{center}
\epsfbox{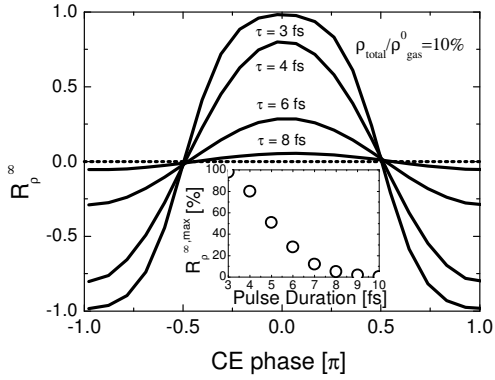} \caption{Asymmetry parameter $R^\infty_\rho$
of the plasma density vs. CE phase for several pulse durations.
$R^\infty_\rho$ is proportional to the macroscopic quasi-DC dipole
moment which is responsible for the emission of the THz radiation.
The pulse intensity is chosen such that a plasma density
$\rho_{total}$ of 10\% of the molecule density $\rho_{gas}^0$is
obtained. Inset: Maximal value of $R^\infty_\rho$ at
$\varphi_{CE}=0$ as a function of the pulse duration.}
\label{asym}
\end{center}
\end{figure}

Electrons generated at positive field polarity experience a
displacement in the positive field direction due to acceleration
in the laser field and incoherent scattering at neighbouring
molecules within a few fs. Correspondingly, electrons generated at
negative polarity are displaced in the negative direction. Such a
collective charge displacement after the passage of the pulse
($\lim_{t \rightarrow \infty}(\rho^+(t) -\rho^-(t)) \neq 0$),
creates a quasi-DC macroscopic dipole moment. Because of its
transient character, it acts as the source of an electromagnetic
pulse, whose polarity and field amplitude are determined by the
ratio $R^\infty_\rho = \lim_{t \rightarrow \infty}((\rho^+-\rho^-)
/ (\rho^+ +\rho^-))$, a quantity which is proportional to the
maximal generated dipole moment. A value of
$R^\infty_\rho$\,=\,$\pm1$ corresponds to the highest possible THz
pulse amplitude, with the polarity of the signal depending on the
sign of $R^\infty_\rho$. For $R^\infty_\rho$\,=\,0, THz emission
vanishes. Figure\,\ref{asym} displays $R^\infty_\rho$ as a
function of $\varphi_{CE}$ for laser pulses with Gaussian envelope
of various durations $\tau$. One finds that the magnitude of
$R^\infty_\rho$ is directly related to the CE phase with a nearly
cosine-like dependence. The contrast, however, rapidly vanishes
when the pulse duration rises towards 10\,fs.

Whilst the theory for stereo ATI requires modelling of both the
quiver motion and recollision of electrons, this is not necessary
for THz emission from a plasma generated at atmospheric pressure.
The reason for this is that the coherent motion of the majority of
electrons is terminated by ultrafast scattering before recollision
and the THz signal is proportional to the asymmetry in the
tunnelling process. Therefore the signal in our method is
conceptually and in contrast to stereo ATI directly related to the
CE phase in a way such that a cosine-like pulse ($\varphi_{CE}=0$)
will produce a maximum asymmetry in ionisation rates and also a
maximum THz-signal. The introduced THz method would measure
directly the absolute CE phase, if the plasma would be a point
object. However, the plasma is an extended object\,\cite{actapol}
and we expect a CE phase variation by dispersion during pulse
propagation and a concomitant interference of the THz signals
generated along the beam path. The strength of this effect for the
moment is unknown, but will be investigated in future measurements
and simulations.

\begin{figure}
\begin{center}
\epsfbox{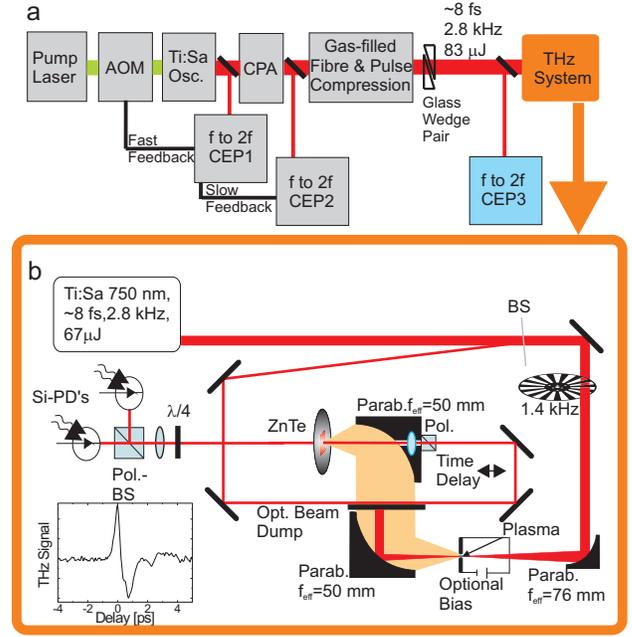} \caption{Sketch of the Experimental set-up.
a) Schematic of the laser system. AOM, acousto-optical modulator;
CEP1, f-to-2f interferometer used for CE phase stabilisation with
fast feedback; CPA, chirped pulse amplification and compression;
CEP2, f-to-2f interferometer used for compensation of slow drifts
of the CE phase; gas-filled fibre used for spectral broadening;
CEP3, f-to-2f interferometer only used for CE phase measurements;
b) THz system employed for electro-optical THz detection. Inset: A
temporal THz wave-form obtained with applied DC bias.} \label{THz}
\end{center}
\end{figure}

\begin{figure}
\begin{center}
\epsfbox{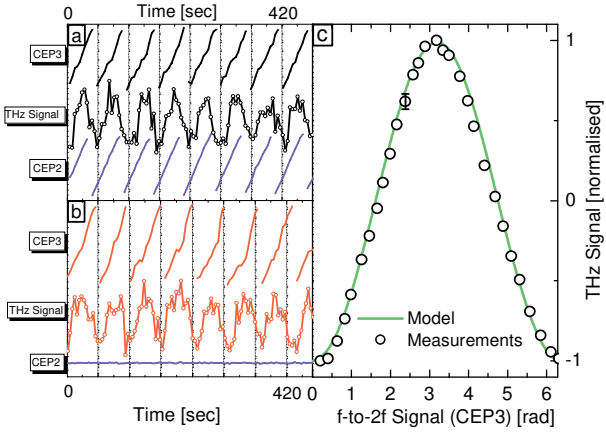} \caption{a) Variation of the CE phase of
8-fs pulses obtained by slowly ramping the set-point of the
control loop. The CE phase is measured by a f-to-2f interferometer
(CEP2) directly behind the laser amplifier, another f-to-2f
interferometer (CEP3) behind the recompressor, and by the THz
measurement system via electro-optical detection of the THz
radiation generated in an ambient air plasma. The integration time
constant of the lock-in amplifier in the THz measurements was
100\,ms. In addition, a sliding average was calculated
numerically, bringing the effective integration time constant to 5
sec. Dispersion effects result in an arbitrary offset between the
CE phases measured by CEP2, CEP3 and the THz system. b) Variation
of the CE phase of 8-fs pulses by a change of the dispersion in
the beam path, realised by two variable glass wedges. c) Adjacent
average of the correlation of the CEP3 and the THz data. The
window chosen for averaging was $\pm$0.5 radians of CEP3. For the
averaging, the total data pool recorded within 74 min. is taken
into account. The error bar depicts the standard deviation of this
averaging. The solid line shows the results of model calculations
of $R^\infty_\rho$ for an 8-fs pulse (constant scaling factor
between theory and experiment). The peak electric field used for
calculations is $E_{opt}^0 = 4\cdot10^{10}$\,V/m corresponding to
the expected value in the experiment.} \label{Comp}
\end{center}
\end{figure}

An experimental test of these theoretical findings was performed
by THz-emission spectroscopy with phase-stabilised 8-fs amplified
laser pulses (see in Methods a detailed description of the set-up
displayed in Fig.\,\ref{THz}). We measured the THz field value
from a laser-generated air plasma at a fixed temporal position
(that with the largest amplitude of the THz transients) while
continuously varying the CE phase by slowly ramping the set-point
of the control loop (with a slope of 2$\pi$ per minute). An
exemplary time series spanning 8 minutes is shown in
Fig.\,\ref{Comp}a. The CE phase data measured in the frequency
domain by two f-to-2f interferometers\cite{baltuska2} (CEP2 and
CEP3) are compared to the THz data. The THz signal varies with the
ramped CE phase, as expected from the model calculations shown in
Fig.\,\ref{asym}.

In a second experiment, the CE phase was kept constant in the
oscillator and laser amplifier (CEP2), but varied with the help of
a piezo-driven glass-wedge pair behind the pulse compressor and in
front of both the third f-to-2f interferometer (CEP3) and the THz
system (also with a phase ramping of $2\pi$ per minute). The
results are depicted in Fig.\,\ref{Comp}b. The measurements again
consolidate the expected change of the THz signal with varying CE
phase.

In order to provide a more quantitative analysis of the
measurements, we show in Fig.\,\ref{Comp}c the THz signal as a
function of the CEP3 data (open dots), averaged over a 74-minute
time series, where we ramped the phase of the oscillator as in
Fig.\,\ref{Comp}a. The error bar indicates the standard deviation
of the THz-signal values. The correlation between THz signal and
CEP3 compares well with the dependence predicted in
Fig.\,\ref{asym} for 8-fs pulses (full line in Fig.\,\ref{Comp}c).
The slight deviation of the theoretical data from the experimental
ones may be a consequence of a non-Gaussian shape of the envelope.

The status of the THz approach for CE phase determination can be
considered as a proof-of-principle. Nevertheless, to classify its
current precision, we compare it to stereo ATI, which achieves a
3$\sigma$ error of 100\,mrad for 10 seconds of data acquisition
and for 6.5-fs 20-$\mu$J pulses\,\cite{paulus2}. The precision
reached with the THz system is 700\,mrad for 10 seconds of data
acquisition for 8-fs 66-$\mu$J pulses. Considering the factor-of-4
improvement to be achieved with 6.5-fs instead of 8-fs pulses (see
Fig.\,\ref{asym}), the THz approach compares well with the much
more mature ATI method already now. The sensitivity of
low-repetition-rate THz-detection can be enhanced significantly by
box-car techniques or by THz interferometry with intensity
detection over a large THz-photon bandwidth.

In summary, we report (i) frequency down-conversion in an
isotropic medium without breaking the symmetry of the non-linear
medium by external means, and (ii) CE phase measurement of
sub-10-fs amplified optical pulses, both by detection of the THz
radiation emitted from a laser-generated ambient-air plasma. The
theory of tunnelling ionisation explains the findings well, in
particular the down-conversion (THz generation) by a spatially
asymmetric process of ionisation and charge acceleration.
THz-emission spectroscopy possesses the potential to become a
commonly employed tool for determination and long term
stabilisation of the CE phase for amplified few-cycle pulses.

\section{METHODS}
\label{meth} \textbf{Ionisation model employed in the
simulations} \\
We base the quantitative description of the photo-ionisation
process of atoms and molecules on static-tunnelling theory
\cite{Corkum89}, with the following expression for the ionisation
rate:

\begin{eqnarray}
\label{eqn1}
\label{w_tun} w_{tun} & = & 4 \omega_a
\left(\frac{U_{ion}^{N_2}}{U_{ion}^{H}}\right)^\frac{5}{2}
\left(\frac{E_{a}}{E_{opt}}\right) \nonumber \\
& & \times \exp \left(-\frac{2}{3}
\left(\frac{U_{ion}^{N_2}}{U_{ion}^{H}}\right)^\frac{3}{2}
\left(\frac{E_{a}}{E_{opt}}\right) \right),
\end{eqnarray}
where $\omega_a$ = 4.13$\cdot$10$^{16}$ 1/s is the atomic
frequency unit, $E_{a}$ = 5.145$\cdot$10$^{11}$ V/m the atomic
unit of the electric field, and $E_{opt}$ the optical field
strength. $U_{ion}^{H}$ = 13.6 eV and $U_{ion}^{N_2}$ = 15.6 eV
denote the ionisation potentials of atomic hydrogen and molecular
nitrogen, respectively. The simple tunnelling model is known to
describe the experimental data for N$_2$ well if the intensity of
the laser pulses is above 10$^{14}$ W/cm$^2$, and to achieve a
fair agreement down to 5$\cdot$10$^{13}$
W/cm$^2$\,\,\cite{Gibson}.

\textbf{Temporal evolution of the ionisation} \\
In order to determine the temporal evolution of the plasma
current, we compute the plasma densities $\rho^+(t)$ and
$\rho^-(t)$ by solving the following coupled
differential equations:
\begin{equation}
\label{diff_ion-1} \frac{\mbox{d}}{\mbox{d}t}\rho^+(t)= \left\{
\begin{array}{ll}
w_{tun}(E_{opt}(t))\cdot(\rho_{gas}^0-\rho^+(t)-\rho^-(t)), & \\
\hspace{1cm} \mbox{ if }E_{opt}(t)\geq0 & \\
0, \hspace{7mm} \mbox{ if }E_{opt}(t)<0 \, ,& \\
\end{array}
\right.
\end{equation}
\begin{equation}
\label{diff_ion-2} \frac{\mbox{d}}{\mbox{d}t}\rho^-(t)= \left\{
\begin{array}{ll}
w_{tun}(E_{opt}(t))\cdot(\rho_{gas}^0-\rho^+(t)-\rho^-(t)), & \\
\hspace{1cm} \mbox{ if }E_{opt}(t)<0 & \\
0, \hspace{7mm} \mbox{ if }E_{opt}(t)\geq0 \, .& \\
\end{array}
\right.
\end{equation}
Here, $\rho_{gas}^0 = 2.7\cdot10^{19}$\,cm$^{-3}$ is the density
of N$_2$ molecules at atmospheric pressure assuming air to only
consist of N$_2$. $E_{opt}(t)$ denotes the electric field strength
of the laser pulse. We neglect plasma recombination which occurs
on a much longer time scale \cite{Biondi}. We assume the following
temporal wave-form of the laser pulse:
\begin{equation}
\label{pulse}
 E_{opt}(t)=E^0_{opt}A(t)\cos\left(\frac{2\pi
c}{\lambda} t + \varphi_{CE}\right),
\end{equation}
where $A(t)$ is a Gaussian envelope. Whilst such a product
separation of carrier and envelope can become ill-defined as the
pulse duration approaches the sub-cycle range\,\cite{Shvartsburg},
it is perfectly reasonable for the shortest pulse durations we
consider here.

\textbf{Light source} \\
The experiments were performed with a light source as sketched in
Fig.\,\ref{THz}a.  A commercial available Ti:Saphire Multipass CPA
System ``Femtopower Compact Pro'' (Femtolasers, Vienna) is seeded
with a self-build laser oscillator (pulse energy 4 nJ, pulse
duration 20fs, centre wavelength 810 nm, CE phase-stabilised with
f-to-2f self referencing\cite{Baltuska} (CEP1)) and driven at a
repetition rate of 2.8 kHz. Its output (500 $\mu$J, 30fs, @805 nm)
is focused into a gas-filled hollow fibre (inner diameter 200
$\mu$m, length 82 cm, 4.5 bar Neon). A small fraction of the CPA
output (few $\mu$J) are used for a f-to-2f
interferometer\cite{baltuska2} (CEP2) to correct for slow CE phase
drifts in the amplifier. The fibre output is then compressed to
its bandwidth limited duration of 8 fs by the use of dispersive
mirrors and a prism-compressor sequence with two fused silica
prisms, with final output parameters (83\,$\mu$J, 8 fs, @750 nm
(centre wavelength)).

\textbf{THz-emission spectroscopy system} \\
80\% of the final output is fed into the THz detection set-up,
which is sketched in Fig.\,\ref{THz}b. 10\% (the residual 10\% are
unused) is used for a third f-to-2f interferometer\cite{baltuska2}
(CEP3), which is tracing CE phase changes for direct comparison
with the THz data in respect to an constant arbitrary offset. The
THz system is a typical THz pump-probe measurement system with
balanced electro-optical THz detection \cite{Löffler,Kress}, the
whole set-up is fitting on a 30$\times$45-cm$^2$ breadboard. The
pump beam is mechanically chopped at half the laser repetition
rate. The plasma is generated in the focus of an off-axis
paraboloidal mirror with an effective focal length of 76 mm. For
optimisation of the alignment and a first determination of the
zero-delay point for the THz detection, an additional DC bias
field can be applied to the focal region \cite{Löffler}. The probe
beam passes along a variable delay line and is focused collinearly
with the THz beam into the electro-optical ZnTe detector crystal.
The THz wave-form as measured with a DC-biased air plasma is
depicted in the lower left part of Fig.\,\ref{THz}b. For the CE
phase measurements, the delay stage of the probe beam is
positioned at the zero-delay point where the strongest THz signal
is obtained, and the external bias field is switched off.



\textbf{Acknowledgements}

We would like to acknowledge fruitful discussions with M.
Horbatsch from York University, Canada. This work was supported in
part by the Hochschulf\"orderungsprogramm of GSI Darmstadt and by
the Deutsche Forschungsgemeinschaft within the contract Mo850/2
(U.M.). Support from the Leibniz-Program of the Deutsche
Forschungsgemeinschaft (J.U.) is gratefully acknowledged.\\

\textbf{Competing interests statement}

The authors declare that they have no competing financial
interests.

\newpage


\end{document}